\documentclass[pdftex]{article}
\usepackage{icrctc07}

\title{Laser Atmospheric Studies with VERITAS}
\shorttitle{Laser Atmospheric Studies with VERITAS}
\authors{C.M.Hui$^{1}$ for the VERITAS collaboration$^{2}$}
\shortauthors{C.M.Hui and et al.}
\afiliations{$^1$University Of Utah, Department of Physics, Salt Lake City,
  UT 84112\\ $^2$ For full author list see G. Maier, ``VERITAS: Status and 
Latest Results'', these proceedings.}
\email{cmhui@physics.utah.edu}

\abstract{As a calibrated laser pulse propagates through the atmosphere,
 the amount of Rayleigh-scattered light arriving at the VERITAS telescopes
 can be calculated precisely.  This technique was originally developed for 
the absolute calibration of ultra-high-energy cosmic-ray fluorescence 
telescopes but is also applicable to imaging atmospheric Cherenkov 
telescopes (IACTs) \cite{ref1}.  In this paper, we present two nights of 
laser data taken with the laser at various distances away from the VERITAS 
telescopes and compare it to Rayleigh scattering simulations.}

\begin{document}
\maketitle
\section{Introduction}
VERITAS, the Very Energetic Radiation Imaging Telescope Array System, is a
GeV-TeV gamma-ray telescope array located at the Fred Lawrence Whipple 
Observatory on Mount Hopkins in Southern Arizona.  It is an array of four 
12m reflectors arranged in a slanted trapezoid with baselines ranging from 
35m to 109m.  Each telescope has a camera comprising 499 photomultiplier 
tubes (PMTs) arranged in a hexagonal lattice and a field of view of 
$3.5^\circ$.  The PMTs are read out via flash-ADCs (FADCs) at a rate of 500 
Msamples/s.  For details see \cite{ref2}.

Standard calibration of IACTs is usually done through measurements of 
the efficiency and gain of individual elements such as electronics and PMTs, 
and through muon-ring measurements.  VERITAS also uses diffused laser
light for relative calibration \cite{ref3}.  However, scattered light from a 
calibrated laser pulse arriving at the telescopes can be detected and simulated 
accurately, allowing absolute calibration to be reliable and simple.

\begin{figure}
\begin{center}
\includegraphics[scale=0.2]{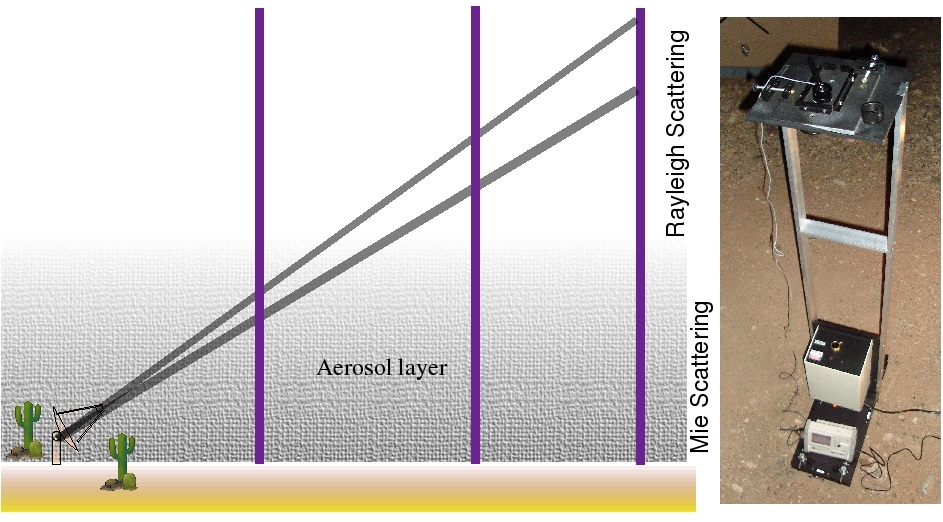}
\end{center}
\caption{The nitrogen laser is fired pointing at zenith with the telescopes
pointed at $20^\circ$ from the horizon.  The laser goes through a layer of aerosol 
and undergoes Mie and Rayleigh scattering.  A picture of the movable laser is shown
on the right.}\label{fig1}
\end{figure}

The nitrogen laser used has a wavelength of 337nm and is mounted on a movable
rack for easy transport.  It is also equipped with flexible beam collimation 
and intensity adjustment. The laser is fired pointing at zenith with the 
telescopes pointing at $20^\circ$ in elevation.  As the laser shot travels
 upward through the atmosphere, the laser light undergoes Rayleigh and Mie 
scattering, some of which is detected by the telescopes.  Depending on 
how far away the laser is fired, the telescopes may intercept the scattered 
light from an altitude that is within the aerosol layer.  Inside the aerosol 
layer, the amount of scattered light increases due to Mie scattering.  Mie 
scattering might even dominate over Rayleigh scattering at these close distances 
(see figure \ref{fig1}).  With measurements at different distances, we can compare 
them to the Rayleigh-scattering simulation and determine how thick and dense the 
aerosol layer was on that particular night.  Furthermore, we can conclude on the 
absolute calibration of individual telescopes once enough data over several nights 
has been collected.

\section{Observations}
The laser measurements presented here were taken in fall 2006, when only two
 of the four telescopes were installed and operating.  On Oct 22 and Nov 24 
of 2006, we took the nitrogen laser to distances between 1 and 7 km away 
from the telescopes.  Both the laser and the telescopes were forced to trigger 
synchronously by two external GPS clocks such that all recorded events contain the 
laser shot.  At the beginning and the end of each 5-minute run, we recorded the ground 
temperature and pressure at the laser firing site for Rayleigh simulation purposes.

Data acquisition of the telescopes was tuned to record 244 FADC samples 
(488 ns) so that each event records the laser shot for as long as the data 
acquisition would allow.  As the laser fires at farther distances, the 
telescopes look at a longer section of the laser beam (see figure \ref{fig1}).
  Due to the limit of recording time in data acquisition and geometric effects,
 the number of pixels that recorded the laser decreases with distance.

\begin{figure*}[h]
\begin{center}
\includegraphics[scale=0.3]{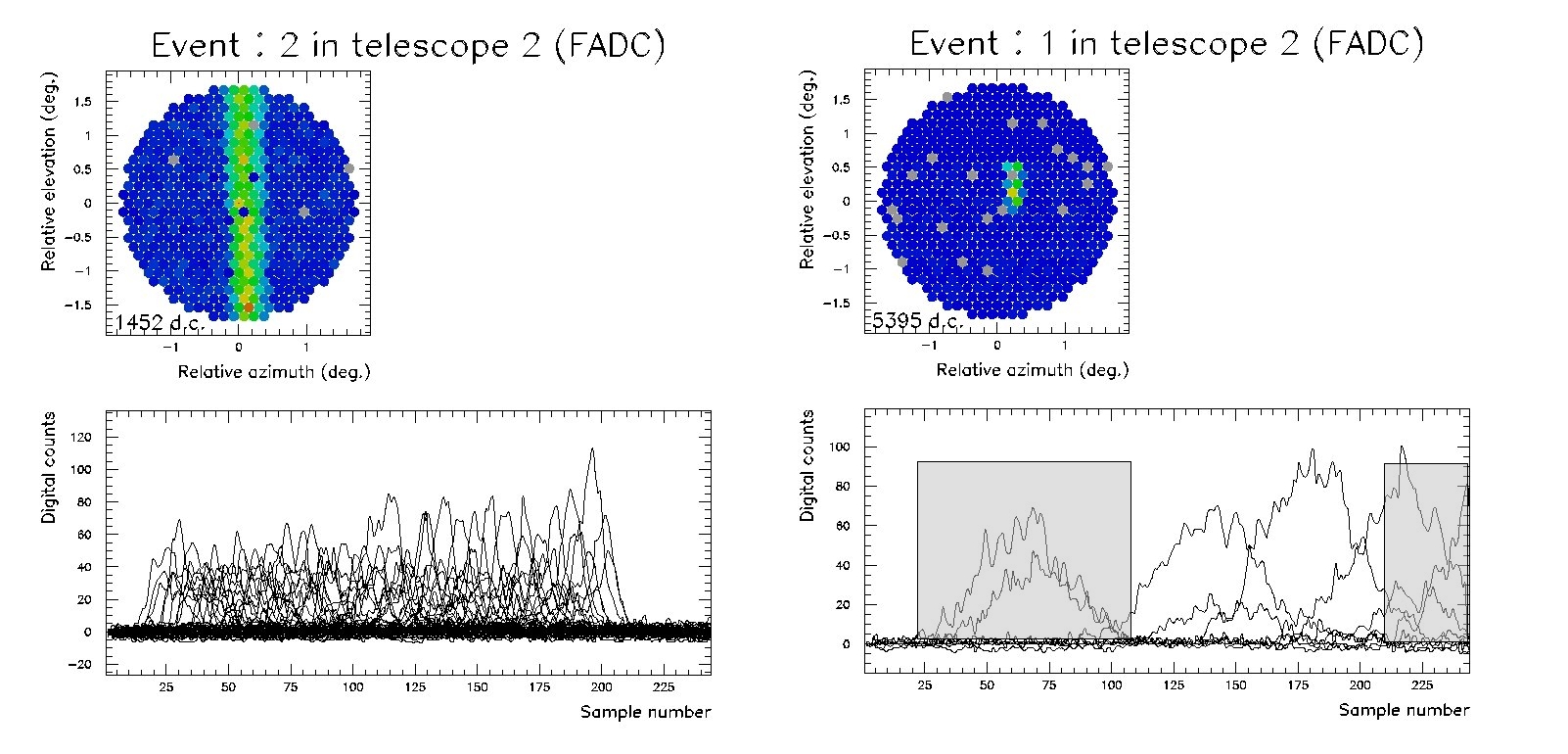}
\end{center}
\caption{On the left shows a sample trace from 1 km away from the telescopes; 
on the right, at 6 km away.  The number of usable pixels decreases with distance
due to data acquisition limit and geometric effects.  At farther distances, the 
telescopes intercept a longer section of the laser beam and each pixel records a
longer laser pulse.  The shaded box is a visual of the box-like function we 
convolved the trace with.  At the far right of the 6km trace, a truncated pulse 
that did not pass our cuts is shown.}\label{fig2}
\end{figure*}

\section{Analysis}
Using an analysis package similar to those described in \cite{ref4}, we convolved 
the FADC trace of each pixel with a box-like function to sum the laser pulse.  The 
width of the box function is adjustable to ensure the entire laser pulse is summed 
since the width of the laser pulse increases with distance (see figure \ref{fig2}).  
The telescopes intercept a longer section of the laser beam at farther distances, 
hence each pixel records a longer laser pulse.  The rest of the samples are averaged
 and used for pedestal subtraction.  To avoid using a truncated pulse, the peak of 
each pulse used in our analysis has to be at least half of the box function width 
away from the first and last recorded samples.  After the cutoffs and the convolution
 are done, we apply a gain-matching factor to the total signal of each pixel such 
that the laser pulse recorded by the pixels is uniform.

\begin{figure*}[h]
\begin{center}
\includegraphics[scale=0.3]{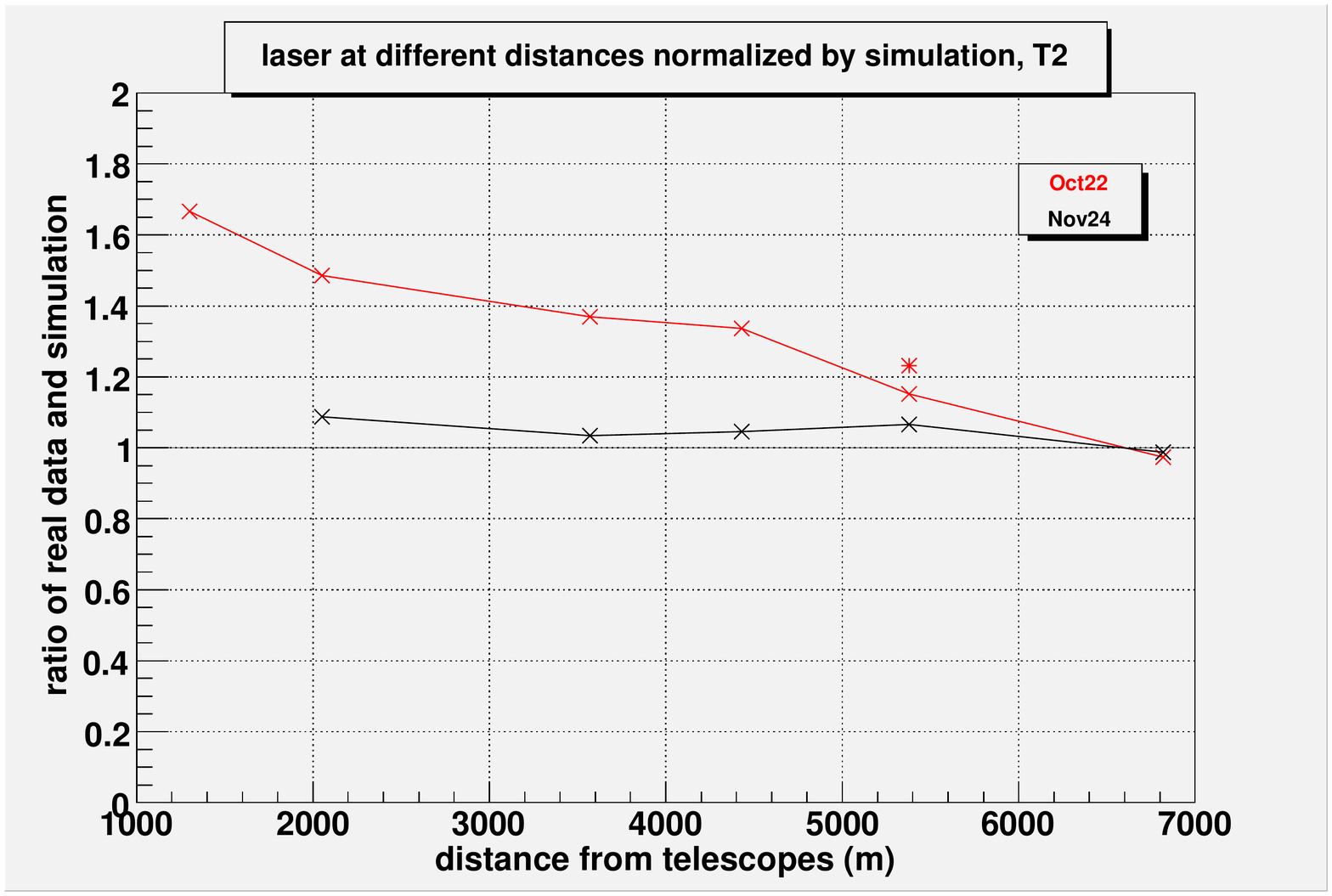}
\end{center}
\caption{Ratio of real data and Rayleigh-scattering simulation of telescope 2 versus
distance between laser and telescope 2.  The $\ast$ indicates an additional data point
 taken at 5 km with a different laser intensity than the original October data set.  
The lighter line shows data from Oct 22, 2006 and the black line shows data from Nov 24,
 2006.  Error bars are not included because there are numerous possible sources from 
the telescopes system that are difficult to quantify.}\label{fig3}
\end{figure*}

Simulations of the VERITAS telescopes' response \cite{ref5} were performed 
together with the Rayleigh-scattering simulation program \cite{ref1}, and 
the same analysis described above was applied to the simulated response.  Figure 
\ref{fig3} displays the ratio between the real data and the Rayleigh-scattering 
simulation.

The gain in the simulation is from the single-photoelectron measurement described in 
\cite{ref3}.  We adjusted the telescopes light collection efficiency such that the 
Rayleigh simulation matches the real data where Rayleigh scattering dominates.  This 
provides the absolute calibration of telescopes.  In the November data set the ratio 
of real data and simulation is close to 1 in all distances, indicating our Rayleigh 
simulation matches the real data closely at all distances and minimal amount of Mie 
scattering occured.  In October, the curve steepens at closer distances, suggesting 
a thicker aerosol layer than in November.

\section{Conclusion}
At closer distances, the telescopes intercept the laser beam inside the aerosol layer 
and were more strongly affected by Mie scattering.  At farther distances, the 
telescopes intercept the laser beam above the aerosol layer and Rayleigh scattering 
dominates.  As shown in figure 3, the amount of light received from the laser at 1 km 
away is nearly doubled of the Rayleigh simulation, whereas at farther distances, the 
Rayleigh simulation matches the real data and the curve flattens.  The thickness of the
aerosol layer changes the distance where the curve plateaus and how steep the curve gets 
at close distances.  With enough data sets from farther distances where Rayleigh 
simulation matches real data, the simulation parameters used could determine the 
absolute calibration of individual telescopes, which could then be used to reduce 
systematic errors in the energy reconstruction of gamma-ray showers.

The aerosol layer changes every night and attenuates Cherenkov showers at an unknown
 level.  A laser study could help determine the nightly aerosol attenuation by 
having a laser setup at a close location and at a distant location.  Taking a set of
 laser measurements prior to observing, the data-simulation comparison curve could 
give information on the aerosol attenuation factor, which could then be applied to 
analysis.

The data acquisition setup for these laser measurements greatly limits the pixel 
statistics at farther distances.  At 1 km away from the telescopes, over 50 pixels 
passed our truncation cut, while at 6 km away, less than 10 pixels passed the cuts.  
The systematic uncertainty in farther measurements are much greater than in closer 
measurements.  The next step is to configure the pixels to read at different memory 
depths of the recorded trace such that all pixels aligned with the image of the scattered 
beam will contain the laser pulse and become usable, alleviating our pixel-statistic 
problem.

\subsection*{Acknowledgments}
This research is supported by grants from the U.S. Department of Energy, the U.S. National 
Science Foundation and the Smithsonian Institution, by NSERC in Canada, by PPARC in 
the U.K. and by Science Foundation Ireland .

\bibliography{icrc0794}

\begin{thebibliography}{1}

\bibitem{ref4}
M.~{Daniel et al.}
\newblock {The VERITAS Standard Data Analysis}.
\newblock In {\em 30th ICRC, Merida}, 2007.

\bibitem{ref3}
D.~{Hanna et al.}
\newblock {Calibration techniques for VERITAS}.
\newblock In {\em 30th ICRC, Merida}, 2007.

\bibitem{ref5}
G.~{Maier et al.}
\newblock {Monte Carlo studies of VERITAS}.
\newblock In {\em 30th ICRC, Merida}, 2007.

\bibitem{ref2}
G.~{Maier et al.}
\newblock {VERITAS: Status and Latest Results}.
\newblock In {\em 30th ICRC, Merida}, 2007.

\bibitem{ref1}
N.~{Shepherd et al}.
\newblock {Absolute calibration of imaging atmospheric Cherenkov telescopes}.
\newblock In {\em 29th ICRC, Pune}, pages 427--430, 2005.

\end{thebibliography}
\bibliographystyle{plain}
\end{document}